\documentclass[showpacs, preprintnumbers, twocolumn, prl]{revtex4}
\usepackage{amssymb}

\usepackage{amsmath}
\usepackage[dvips]{graphicx}

\begin{document}

\title{Generation of macroscopic pair-correlated atomic beams by four-wave mixing
in Bose-Einstein condensates}

\begin{abstract}
By colliding two Bose-Einstein condensates we have observed strong bosonic
stimulation of the elastic scattering process. When a weak input beam was
applied as a seed, it was amplified by a factor of 20. This large gain
atomic four-wave mixing resulted in the generation of two macroscopically
occupied pair-correlated atomic beams.

\pacs{05.30.Jp, 03.75.Fi, 32.80.-t, 67.40.Db}%
%
\end{abstract}

\author{J.M. Vogels}
\author{K. Xu}
\author{W. Ketterle}
\homepage[Group website: ]{http://cua.mit.edu/ketterle_group/}
\affiliation{Department of Physics, MIT-Harvard Center for Ultracold
Atoms, and Research Laboratory of Electronics,
Massachusetts Institute of Technology, Cambridge, MA 02139}
\maketitle%
%
\catcode`\ä = \active \catcode`\ö = \active \catcode`\ü = \active
\catcode`\Ä = \active \catcode`\Ö = \active \catcode`\Ü = \active
\catcode`\ß = \active \catcode`\é = \active \catcode`\è = \active
\catcode`\ë = \active \catcode`\ô = \active \catcode`\ê = \active
\catcode`\ø = \active \catcode`\ò = \active \catcode`\í = \active
\defä{\"a} \defö{\"o} \defü{\"u} \defÄ{\"A} \defÖ{\"O} \defÜ{\"U} \defß{\ss} \defé{\'{e}}
\defè{\`{e}} \defë{\"{e}} \defô{\^{o}} \defê{\^{e}} \defø{\o} \defò{\`{o}} \defí{\'{i}}%
%
Bose-Einstein condensation (BEC) in a dilute atomic gas produced a
macroscopic population of the ground state wavefunction \cite{ingu99}. Once
BEC had been achieved, the initial well-defined quantum state can be
transformed into other more complex states by manipulating it with magnetic
and optical fields. This can result in a variety of time-dependent
macroscopic wavefunctions \cite{ingu99}, including oscillating condensates,
multiple condensates moving relative to each other, an output coupler and
rotating condensates with vortex lattices. Such macroscopically occupied
quantum states represent classical matter-wave fields in the same way an
optical laser beam is a classical electromagnetic wave. The next major step
involves engineering non-classical states of atoms that feature quantum
entanglement and correlations. These states are important for quantum
information processing, sub-shot noise precision measurements \cite{orze01},
and tests of quantum non-locality.

Quantum correlations in the BEC ground state have been observed in a BEC
held in optical lattices \cite{Jaks99,orze01,Grei2002}. The repulsive
interactions between the atoms within each lattice site forces the
occupation numbers to equalize, resulting in a number squeezed state.
Alternatively, correlations in a BEC can be created in a dynamic or
transient way through interatomic collisions. At the low densities typical
of current experiments, binary collisions dominate, creating correlated
pairs of atoms. Due to momentum conservation, the pair-correlated atoms
scatter into modes with opposite momenta in the center-of-mass frame,
resulting in squeezing of the number difference between these modes \cite
{Duan2000,Pu2000,Sore2001}. Our work is an implementation of the suggestions
in Refs.\ \cite{Duan2000,Pu2000}. However, we use elastic scattering
processes instead of spin flip collisions to create pair correlations
because the elastic collision rate is much higher than the spin flip rate.
This was essential to observe large amplification before further elastic
collisions led to losses.

Elastic scattering between two BEC's produces a collisional halo \cite
{chik00}, where the number of atoms moving into opposing solid angles is the
same, corresponding to number squeezing. Once these modes are occupied, the
scattering process is further enhanced by bosonic stimulation. The onset of
such an enhancement was observed in Ref.\ \cite{chik00}. In this paper, we
report strong amplification, corresponding to a gain of at least 20. Based
on a theoretical prediction  \cite{Pu2000}, which drew an analogy to optical
superradiance \cite{inou99super}, we expected to obtain a highly anisotropic
gain using our cigar-shaped condensate. However, this mechanism of mode
selection proved to be irrelevant for our experiment, because atoms do not
leave the condensate during amplification (see below). Instead, we preselect
a single pair-correlated mode by seeding it with a weak third matter wave,
and observed that up to 40 \% of the atoms scattered into it. Because the
scattered atoms are perfectly pair-correlated, the only fluctuations in the
number difference between the two beams stem from number fluctuations in the
initial seed. Therefore, an observed gain of 20 implies that we have
improved upon the shot-noise limit by a factor of $\sqrt{20}$, although this
was not directly observed. Such a four-wave mixing process with matter waves
had only been observed previously with a gain of 1.5 \cite{deng99,Trip2000}.

\begin{figure}[tbp]
\includegraphics[width=85mm]{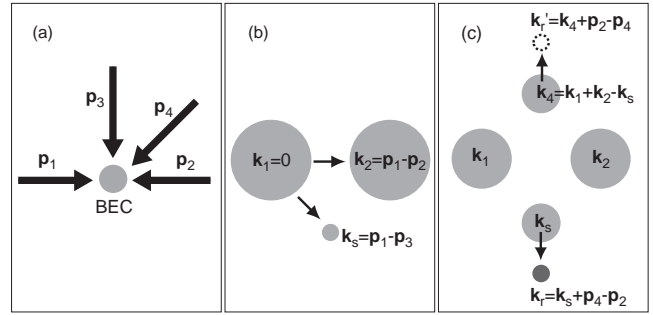}
\caption{Arrangement of laser beams to generate three atomic wavepackets:
(a) Four Bragg beams intersected at the condensate. (b) Two source waves and
a small seed were created with Bragg beam pairs. (c) The four-wave mixing
process amplified the seed and created a fourth wave. Both were subsequently
read out with another Bragg pulse. The figures are projections on a plane
perpendicular to the condensate axis. All wavepackets move within this plane.
}
\label{fig:scheme}
\end{figure}

This experiment was performed with sodium condensates of $\sim $30 million
atoms in a cigar-shaped magnetic trap with radial and axial trap frequencies
of 80 Hz and 20 Hz, respectively. Condensates had a mean field energy of $4.4
$ kHz, a speed of sound of 9 mm/s, and radial and axial Thomas-Fermi radii
of 25 $\mu $m and 100 $\mu $m, respectively. The second condensate and the
seed wave were generated by optical Bragg transitions to other momentum
states. Fig. \ref{fig:scheme} shows the geometry of the Bragg beams. Four
laser beams were derived from the same laser that was 100 GHz red-detuned
from the sodium $D_{2}$ line. The large detuning prevented optical
superradiance \cite{inou99super}. All beams propagated at approximately the
same angle of $\sim $ 0.5 rad with respect to the long axis of the
condensate, and could be individually switched on and off to form beam pairs
to excite two-photon Bragg transitions \cite{sten99brag} at different recoil
momenta.

\begin{figure}[tbp]
\includegraphics[width=85mm]{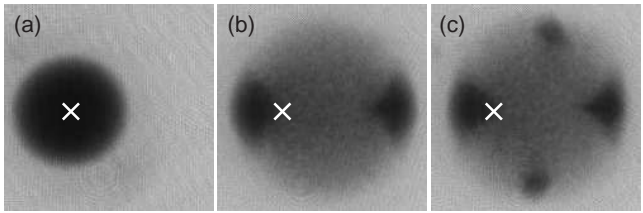}
\caption{High-gain four wave mixing of matter waves. The wavepackets
separated during 43 ms of ballistic expansion. The absorption images 
\protect\cite{ingu99} were taken along the axis of the condensate. (a) Only
a 1\% seed was present (barely visible), (b) only two source waves were
created and no seed, (c) two source waves and the seed underwent the
four-wave mixing process where the seed wave and the fourth wave grew to a
size comparable to the source waves. The gray circular background consists
of spontaneously emitted atom pairs that were subsequently amplified to
around 20 atoms per mode. The crosses mark the center position of the
unperturbed condensate. The field of view is 1.8 mm wide.}
\label{fig:growth}
\end{figure}

The seed wave was created by a weak 20 $\mu $s Bragg pulse of beams with
momenta $\mathbf{p}_{1}$ and $\mathbf{p}_{3}$, which coupled 1-2 \% of the
atoms into the momentum state $\mathbf{k}_{s}=\mathbf{p}_{1}-\mathbf{p}_{3}$%
, with a velocity $\sim $15 mm/s. Subsequently, a 40 $\mu $s $\pi /2$ pulse
of beams $\mathbf{p}_{1}$ and $\mathbf{p}_{2}$ split the condensate into two
strong source waves with momenta $\mathbf{k}_{1}=0$ and $\mathbf{k}_{2}=%
\mathbf{p}_{1}-\mathbf{p}_{2}$ (see Fig.~\ref{fig:scheme}b), corresponding
to a relative velocity of $\sim $20 mm/s. The four wave mixing process
involving these three waves led to an exponential growth of the seed wave
while a fourth conjugate wave at momentum $\mathbf{k}_{4}=\mathbf{k}_{1}+%
\mathbf{k}_{2}-\mathbf{k}_{s}$ emerged and also grew exponentially (Fig.~\ref
{fig:scheme}c). The Bragg beams were arranged in such a way that the phase
matching condition was fulfilled (the sum of the kinetic energies of the
source waves ($\sim 11$ kHz) matched the energy of the seed and the fourth
wave). The effect of any energy mismatch on the process will be discussed
later. The four-wave mixing process was analyzed by absorption imaging\cite
{ingu99}. Fig.~\ref{fig:growth}c shows the key result of this paper
qualitatively: A small seed and its conjugate wave were amplified to a size
where a significant fraction of the initial condensate atoms had been
transferred into this pair-correlated mode.

\begin{figure}[tbp]
\includegraphics[width=75mm]{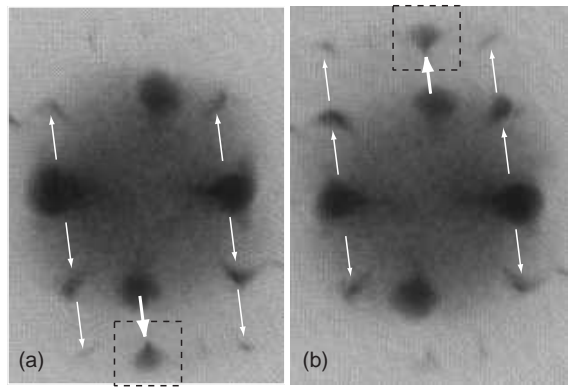}
\caption{Absorption images after a readout pulse was applied to (a) the seed
wave, and (b) the fourth wave. The thick arrows indicate the readout
process. The readout pulse was kept short (40 $\protect\mu $s), resulting in
a large Fourier bandwidth and off-resonant coupling to other wave packets
indicated by the narrow arrows. However, this did not affect the readout
signal (atoms in the dashed box).}
\label{fig:readout}
\end{figure}

To study this process, we applied ``readout'' beams $\mathbf{p}_{2}$ and $%
\mathbf{p}_{4}$ for 40 $\mu $s, interrupting the amplification after a
variable growth period between 0 and 600 $\mu $s. (Turning off the trap
after a variable amount of time is insufficient in this case because the
density decreases on the time scale of the trapping period (1/(80 Hz)),
while the amplification occurs more rapidly.) The frequency difference
between the two readout beams was selected such that a fixed fraction of the
seed (fourth) wave was coupled out to a different momentum state $\mathbf{k}%
_{r}=\mathbf{k}_{s}+\mathbf{p}_{4}-\mathbf{p}_{2}$ ($\mathbf{k}_{r}^{\prime
}=\mathbf{k}_{4}+\mathbf{p}_{2}-\mathbf{p}_{4}$) (Fig.~\ref{fig:scheme}c). $%
\mathbf{k}_{r}$ or $\mathbf{k}_{r}^{\prime }$ did not experience further
amplification due to the constraint of energy conservation and therefore
could be used to monitor the atom number in the seed (fourth) wave during
the four-wave mixing process (Fig.~\ref{fig:readout}).

\begin{figure}[tbp]
\includegraphics[width=85mm]{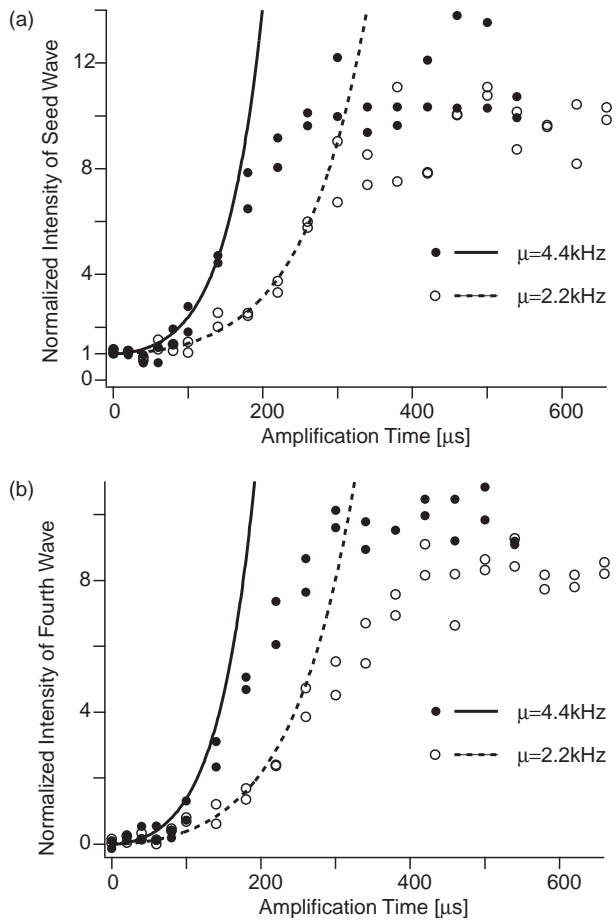}
\caption{Generation of pair-correlated atomic beams. The growth of (a) a 2
\% seed and (b) it's conjugate fourth wave are shown for two different
chemical potentials $\protect\mu $. The intensity of the waves are
normalized to the intensity of the initial seed. The solid and dashed lines
follow the initial growth according to Eq. (\ref{eq3}) with growth rates of
(170 $\protect\mu $s)$^{-1}$ and (100 $\protect\mu $s)$^{-1}$, respectively.}
\label{fig:data}
\end{figure}

The growth of the 2\ \% seed and the fourth wave are shown in Fig.~\ref
{fig:data}. As expected, the growth rates were found to increase with the
mean field energy. Eventually, the amplification slowed down and stopped as
the source waves were depleted. This is in contrast to Ref.~\cite{deng99},
where the mixing process was slow (due to much lower mean field energy), and
the growth time was limited by the overlap time of the wavepackets. In our
experiment, the overlap time was $\gtrsim 1.8$ ms, whereas the growth
stopped already after $\lesssim 500$ $\mu $s.

A simple model describes the salient features of the process. The
Hamiltonian of a weakly interacting Bose condensate is given by \cite
{Trip2000} 
\begin{equation}
H=\sum\limits_{\mathbf{\kappa }}\frac{\hbar ^{2}\mathbf{\kappa }^{2}}{2m}%
\widehat{a}_{\mathbf{\kappa }}^{\dagger }\widehat{a}_{\mathbf{\kappa }}+%
\frac{2\pi \hbar ^{2}a}{mV}\sum_{\substack{ \mathbf{\kappa }_{1}+\mathbf{%
\kappa }_{2} \\ =\mathbf{\kappa }_{3}+\mathbf{\kappa }_{4}}}\widehat{a}_{%
\mathbf{\kappa }_{1}}^{\dagger }\widehat{a}_{\mathbf{\kappa }_{2}}^{\dagger }%
\widehat{a}_{\mathbf{\kappa }_{3}}\widehat{a}_{\mathbf{\kappa }_{4}}
\label{eq1}
\end{equation}
where $\mathbf{\kappa }$ denotes the wavevectors of the plane wave states, $m
$ is the mass, $V$ is the quantization volume, $\widehat{a}_{\mathbf{\kappa }%
_{i}}$ is the annihilation operator and $a=2.75$ nm is the scattering
length. If two momentum states $\mathbf{k}_{1}$ and $\mathbf{k}_{2}$ are
highly occupied relative to all other states (with occupation numbers $N_{1}$
and $N_{2}$), the initial depletion of $\mathbf{k}_{1}$ and $\mathbf{k}_{2}$
can be neglected. Therefore, the only interactions are mean field
interactions(self-interactions) and scattering involving $\mathbf{k}_{1}$
and $\mathbf{k}_{2}$. In the Heisenberg picture, the difference between the
occupations of the mode pairs $\Delta \widehat{n}=\widehat{a}_{\mathbf{%
\kappa }_{1}}^{\dagger }\widehat{a}_{\mathbf{\kappa }_{1}}-\widehat{a}_{%
\mathbf{\kappa }_{2}}^{\dagger }\widehat{a}_{\mathbf{\kappa }_{2}}$ is time
independent for any $\mathbf{\kappa }_{1}+\mathbf{\kappa }_{2}=\mathbf{k}%
_{1}+\mathbf{k}_{2}$ . Therefore, the fluctuations in the number difference $%
\langle \Delta \widehat{n}^{2}\rangle -\langle \Delta \widehat{n}\rangle ^{2}
$ remain constant even though the occupations grow in time. The result is
two-mode number squeezing. This is equivalent to a non-degenerate parametric
amplifier---the Hamiltonian for both systems are identical \cite{wall1995}.

When calculating the occupations $\langle \widehat{a}_{\mathbf{\kappa }%
_{1}}^{\dagger }\widehat{a}_{\mathbf{\kappa }_{1}}\rangle $, $\langle 
\widehat{a}_{\mathbf{\kappa }_{2}}^{\dagger }\widehat{a}_{\mathbf{\kappa }%
_{2}}\rangle $ and the correlation $\langle \widehat{a}_{\mathbf{\kappa }%
_{1}}\widehat{a}_{\mathbf{\kappa }_{2}}\rangle $, the relevant physical
parameters are
\begin{eqnarray}
\overline{\mu } &=&\sqrt{\mu _{1}\mu _{2}}\text{ \ , \ }\mu _{i}=\frac{4\pi
\hbar ^{2}a}{mV}N_{i}\text{ \ }(i=1,2)  \notag \\
\Delta \omega  &=&\frac{\hbar \mathbf{\kappa }_{1}^{2}}{2m}+\frac{\hbar 
\mathbf{\kappa }_{2}^{2}}{2m}-\frac{\hbar \mathbf{k}_{1}^{2}}{2m}-\frac{%
\hbar \mathbf{k}_{2}^{2}}{2m}+\frac{\mu _{1}}{\hbar }+\frac{\mu _{2}}{\hbar }
\label{eq4}
\end{eqnarray}
where $\overline{\mu }$ is the (geometric) average mean field energy of the
two source waves, and$\ \hbar \Delta \omega $ is the energy mismatch for the
scattering of atoms from states $\mathbf{k}_{1}$ and $\mathbf{k}_{2}$ to
states $\mathbf{\kappa }_{1}$ and $\mathbf{\kappa }_{2}$. One obtains
exponential growth for $\mathbf{\kappa }_{1}$ and $\mathbf{\kappa }_{2}$ if $%
\overline{\mu }>\hbar \Delta \omega /4$, and the growth rate is given by:
\begin{equation}
\eta =\sqrt{\left( \frac{2\overline{\mu }}{\hbar }\right) ^{2}-\left( \frac{%
\Delta \omega }{2}\right) ^{2}}  \label{eq2}
\end{equation}

For our initial conditions with $s$ atoms in the seed wave $\mathbf{k}_{s}$
and an empty fourth wave $\mathbf{k}_{4}$, the correlation $\langle \widehat{%
a}_{\mathbf{k}_{s}}\widehat{a}_{\mathbf{k}_{4}}\rangle $ start to grow as:
\begin{equation*}
|\left\langle \widehat{a}_{\mathbf{k}_{s}}\widehat{a}_{\mathbf{k}%
_{4}}\right\rangle |=\frac{2\overline{\mu }\sqrt{4\overline{\mu }^{2}\cosh
(\eta t)^{2}-\frac{\Delta \omega ^{2}}{4}}}{\eta ^{2}}\sinh \left( \eta
t\right) (s+1)\text{.}
\end{equation*}
This leads to exponential growth of the occupation numbers:
\begin{eqnarray}
\langle \widehat{a}_{\mathbf{k}_{s}}^{\dagger }\widehat{a}_{\mathbf{k}%
_{s}}\rangle  &=&\left[ \frac{2\overline{\mu }}{\hbar \eta }\sinh \left(
\eta t\right) \right] ^{2}(s+1)+s  \notag \\
\langle \widehat{a}_{\mathbf{k}_{4}}^{\dagger }\widehat{a}_{\mathbf{k}%
_{4}}\rangle  &=&\left[ \frac{2\overline{\mu }}{\hbar \eta }\sinh \left(
\eta t\right) \right] ^{2}(s+1)  \label{eq3}
\end{eqnarray}

Eq. (\ref{eq2}) and (\ref{eq3}) show that for a large mean field energy $%
\overline{\mu }$, $\Delta \omega $ can be quite large without suppressing
the four-wave mixing process. When $\Delta \omega >4\overline{\mu }/\hbar $,
one has to replace the hyperbolic sine functions in Eq.\ (\ref{eq3}) with
sine functions and $\eta $ in Eq.\ (\ref{eq2}) with $\sqrt{(\Delta \omega
/2)^{2}-\left( 2\overline{\mu }/\hbar \right) ^{2}}$. The occupations in
states $\mathbf{k}_{s}$ and $\mathbf{k}_{4}$ still grow initially, but then
they begin to oscillate. The above solution also applies to initially empty
modes with $s=0$.

We can estimate the maximum growth rate ($2\overline{\mu }/\hbar $) for our
experiment by using the average mean field energy across the condensate to
obtain (53 $\mu $s)$^{-1}$ for high and (110 $\mu $s)$^{-1}$ for low mean
field energy. The experimental data exhibit a somewhat slower growth rate of
(100 $\mu $s)$^{-1}$ and (170 $\mu $s)$^{-1}$, respectively. This
discrepancy is not surprising since our theoretical model does not take into
account depletion and possible decoherence processes due to the finite size
and inhomogeneity of a magnetically trapped condensate. We also observed
that the angles of the Bragg beams and therefore the energy mismatch $\Delta
\omega $, could be significantly varied without substantially affecting the
four-wave mixing process, confirming the robustness (see Eq.~(\ref{eq2})) of
four-wave mixing.

In addition to the four distinct wavepackets, Fig.~\ref{fig:growth} also
shows a circular background of atoms that are scattered from the source
waves $\mathbf{k}_{1}$ and $\mathbf{k}_{2}$ into other pairs of initially
empty modes $\mathbf{\kappa }_{1}$ and $\mathbf{\kappa }_{2}$ ($\mathbf{\
\kappa }_{1}+\mathbf{\kappa }_{2}=\mathbf{k}_{1}+\mathbf{k}_{2}$). The
scattered atoms lie on a spherical shell in momentum space centered at $(%
\mathbf{k}_{1}+\mathbf{k}_{2})/2$ with a radius $\left| \mathbf{k}\right| $
close to $\left| \mathbf{k}_{1}-\mathbf{k}_{2}\right| /2$ and a width $%
\left| \Delta \mathbf{k}\right| \sim m\sqrt{2\pi \overline{\mu }/t\hbar ^{3}}%
/\left| \mathbf{k}\right| $. As time progresses, the thickness $\left|
\Delta \mathbf{k}\right| $ narrows due to the exponential gain.

Eventually, the population of these background modes contribute to the
depletion of the source waves. One can estimate the depletion time $t_{d}$
of the source waves by comparing the total population in these modes to the
original number of atoms. This sets a theoretical limit on the gain $G=e^{4%
\overline{\mu }t_{d}/\hbar }/4$ given by $4G/\sqrt{\ln (4G)}=\sqrt{2\pi }%
/\left| \mathbf{k}\right| a$. For our geometry $\left| \mathbf{k}\right|
a=0.01$ and $G=160$. In our condensate of $3\times 10^{7}$ atoms, this
maximum gain is achieved when all the atoms are scattered into the $9\times
10^{4}$ pair modes in the momentum shell. With a 1 \% seed, the source waves
are depleted earlier, leading to a maximum gain of 37, comparable to our
measured gain of 20.

In our experiment, we deliberately reduced the velocity between the two
source waves to twice the speed of sound in order to increase $G$ and also
the overlap time between the two source waves. Under these circumstances,
the thickness of the shell $\left| \Delta \mathbf{k}\right| $ becomes close
to its radius, accounting for the uniform background of scattered atoms
rather than the thin $s$-wave halo observed in Ref. \cite{chik00}. For
velocities around or below the speed of sound the condensate won't separate
from the other waves in ballistic expansion.

Once the amplified modes are populated, losses due to further collisions
occur at a rate $\Gamma \sim 8\pi a^{2}n\hbar |\mathbf{k}|/m$ per atom ($n$
is the number density of atoms). In order to have net gain, the growth rate $%
\eta $\ should be greater than $\Gamma $, which is the case since $\eta
/\Gamma =1/|\mathbf{k}|a=100\gg 1$. Furthermore, we begin to lose squeezing
when $s+1$ atoms are lost from the mode pair that occurs approximately at a
gain of $e^{4\overline{\mu }t/\hbar }/4=1/|\mathbf{k}|a$. At this point, the
condensate is already highly depleted. In our experiment however, the shell
of amplified modes is so thick that it includes many of the modes into which
atoms are scattered and increases the scattering rate by bosonic
stimulation. Ideally, the atomic beams should separate after maximum gain is
achieved. However, for our condensate size, the waves overlap for a much
longer time and suffer collisional losses. This is visible in Fig. 2, where
40\% of the atoms were transferred to the seeded mode pair, but only $\sim $
10\% survived the ballistic expansion.

The collisional amplification process studied here bears similarities to the
superradiant Rayleigh scattering of light from a Bose condensate \cite
{inou99super}, where correlated photon-atom pairs are generated in the
end-fire mode for the photons and the corresponding recoil mode for the
atoms. However, there are significant differences between the two processes.
In optical superradiance, the scattered photons leave the condensate very
quickly, causing only the recoiled atoms to maintain the coherence and
undergo exponential growth. This physical situation is reflected in the
Markov approximation adopted in Refs.~\cite{moor99super,Pu2000}. In
contrast, the atoms move slowly in collisional amplification, and the Markov
approximation does not hold (although it was applied in Ref.~\cite{Pu2000}).
The energy uncertainty $\Delta E=\hbar /\Delta t$ for a process of duration $%
\Delta t$ gives a longitudinal momentum width of $\Delta E/v$ where $v$ is
the speed of light for photons or the velocity of the scattered atoms. This
shows that optical superradiance is much more momentum selective --- the
shell in momentum space is infinitesimally thin, and only the atomic modes
with maximal overlap with this shell are selected. In contrast, the shell in
collisional amplification is many modes thick and does not lead to strong
mode selection. Moreover, in optical superradiance the light is coherently
emitted by the entire condensate, whereas collisional amplification reflects
only local properties of the condensate, because the atoms do not move
significantly compared to the size of the condensate. Therefore, features
like growth rate, maximum amplification, and even whether mode pairs stay
squeezed do not depend on global parameters such as size or shape.

In conclusion, we have observed high gain in atomic four-wave mixing and
produced pair-correlated atomic beams. We have also identified some
limitations for using collisions to create such twin beams, including loss
by subsequent collisions, and competition between other modes with similar
gain.

We thank Michael Moore for useful discussions and for sending us early
drafts of his recent theory paper \cite{Vard2002}. The conclusions of his
paper agree with ours. We also thank James Anglin and Peter Zoller for
helpful interactions, Jamil Abo-Shaeer for experimental assistance, and Jit
Kee Chin for critical reading of the manuscript. This work was funded by
ONR, NSF, ARO, NASA, and the David and Lucile Packard Foundation.

\end{document}